\documentclass[twocolumn,twoside,slac_two]{revtex4}
\usepackage{graphicx}
\usepackage{fancyhdr}
\begin{document}

\title{Fermi Bubbles as a Result of Star Capture in the Galactic Center}

%

\author{V.A. Dogiel, D.O. Chernyshov}
\affiliation{I.E.Tamm Theoretical Physics Division of P.N.Lebedev
 Institute of Physics, Leninskii pr. 53, 119991, Moscow, Russia}
\author{K.-S. Cheng, Y. Wang}
\affiliation{Department of Physics, The University of Hong Kong,
Pokfulam Road, Hong Kong, China}
\author{C.-M. Ko, W.-H. Ip}
\affiliation{Institute of Astronomy, National Central University,
Jhongli 320, Taiwan }
\begin{abstract}
Fermi has discovered two giant gamma-ray-emitting bubbles that
extend nearly 10 kpc in diameter. We propose that periodic star
capture processes by the galactic supermassive black hole, Sgr A*,
with a capture rate $<10^{-5}$ yr$^{-1}$ and energy release $\sim
10^{52}$ erg per one capture can produce shocks in the halo, which
accelerate electrons to the energy ~ 1 TeV. These electrons
generate radio emission via synchrotron radiation, and gamma-rays
via inverse Compton scattering with the relic and the galactic
soft photons. Estimates of the diffusion coefficient from the
observed gamma-ray flux explains consistently the necessary
maximum energy of electrons and sharp edges of the bubble.

\end{abstract}

\maketitle

\thispagestyle{fancy}


\section{INTRODUCTION}
The recently discovered Fermi bubbles are symmetric gamma-ray
structures derived from the Fermi LAT data in the energy range
1-100 GeV. The bubbles elongate above and below the Galactic plane
for about 8 kpc and their radius is about 3 kpc . Observations
show a very sharp outer boundary of the bubble. The gamma-ray
intensity sharply drops outward the bubbles \cite{meng}.

The origin of the bubble is still enigmatic and up to now a few
models were presented in the literature. Our group assumed that
the Fermi bubbles originated from star capture events which
occurred in the GC every $10^4-10^5$ years \cite{cheng}. These
events form giant shocks propagating through the central part of
the Galactic halo and thus produce accelerated electrons with
energies $\leq 10$ TeV whose scattering on background photons is
responsible for the bubble gamma-ray emission.

Processes of particle acceleration by the bubble shocks in terms
of sizes of the envelope, maximum energy of accelerated particles,
etc. may differ significantly from those obtained for SNs that may
lead to the maximum energy of accelerated protons much larger than
can be reached in SNRs. In this respect, we assume that
acceleration of protons in Fermi bubbles may contribute to the
total flux of the Galactic cosmic rays (CR) above the 'knee' break
($\geq 10^{15}$ eV).

\section{BUBBLE HYDRODYNAMICS}
We assume in \cite{cheng} that the central black hole captures a
star every $\tau_0\sim 10^4-10^5$ years. As a result the total
energy $\mathcal{E}_0\sim 10^{52}$ erg releases in the Galactic
center in the form of 100 MeV proton which heat the central 20 pc
up to the temperature $\sim 10$ keV. This heating produces a shock
propagating into the surrounding medium. In the simplest case this
situation can be described by a solution obtained by \cite{komp}
for the adiabatic explosion in the exponential atmosphere with the
density profile $\rho(z)$,
\begin{equation}
\rho(z)=\rho_0\exp\left(-\frac{z}{z_0}\right)\,.
\end{equation}
 where $z$ is the coordinate
perpendicular to the Galactic plane. For  parameters of the
Galactic halo $\rho_0=0.25$ cm$^{-3}$ and $z_0=1$ kpc. The shock
propagating in to the halo forms in the exponential atmosphere a
double bubble structure elongated in z-direction. The radius of
the bubble at the height $z$ and at the time $t$ is
\begin{equation}
r=2z_0\arccos\left[\frac{1}{2}e^\frac{z}{2z_0}\left(1-\left(\frac{y}{2z_0}\right)^2+
e^{-\frac{z}{z_0}}\right)\right]\,,\label{r_shock}
\end{equation}
where
\begin{equation}
y=\int\limits^t_0\left(\frac{\gamma^2-1}{2}\lambda\frac{\alpha
\mathcal{E}_0}{V(t)\rho_0}\right)^{0.5}dt\,,
\end{equation}
$V$ is a current volume bounded by the shock
\begin{equation}
V(t)=2\pi\int\limits^{a(t)}_0r^2(z,t)dz\,,
\end{equation}
$a$ is the position of the shock top
\begin{equation}
a(t)=-2z_0\ln\left(1-\frac{y}{2z_0}\right)\,,
\end{equation}
$\gamma$ is the polytropic coefficient, and $\alpha$ and $\lambda$
are numbers.

For the finite time $t_1$ determined from the condition
$y(t_1)=2z_0$ the shock breaks through the exponential atmosphere
and the bubble top $a(t_1)$ tends to infinity  while the bubble
radius in the Galactic plane ($z=0$) tends asymptotically to the
the value
\begin{equation}
r=2z_0\arccos\left({1}/{2}\right)\simeq 2~\mbox{kpc}\,,
\end{equation}
 that is comparable with the radius of Fermi bubbles. For
 $\mathcal{E}_0\sim  10^{52}$ erg and $\rho_0 \sim
0.25$ cm$^{-3}$ the value of $t_1$ is about $3\times 10^8$ yr. We
want to remark that $t_1$ sensitively depends on the injected
energy and the density profile of the halo. For example if the
injected energy is $\mathcal{E}_0\sim 3\times 10^{52}$ erg and
$\rho_0 \sim 0.1$ cm$^{-3}$, $t_1$ can be reduced by nearly an
order of magnitude.

 Then
for a periodic star capture the bubble interior is filled with
shocks propagating in series one after another through the bubble
interior stopping at the radius $\simeq 2z_0$ That gives formally
a stationary sideway boundary.

The realistic situation has to be described by a set of
dissipative hydrodynamic equations, which take into account the
shocks propagation in non-uniform medium and various dissipation
processes including shock heating, energy transfer into cosmic
rays, slowing down due to accumulating material etc. These
processes are ignored in the the Kompaneetz solution. Shocks
should disappear when their speed is lower the local sound speed.
shocks should disappear when their speed is lower the local sound
speed. Then the sideway boundary of the Bubble is simply given by
$r_b \sim v_s t_{dis}$ where $v_s$ is the sound speed and
$t_{dis}$ is the characteristic time of the shock dissipation
because of e.g. particle acceleration at the shock front.
\section{ELECTRON ACCELERATION AND GAMMA-RAYS FROM THE BUBBLE}
We assume that the bubble gamma-rays are produced by IC scattering
of electrons on the relic photons. For the rate of synchrotron and
inverse Compton  energy losses  $dE/dt=\beta E^2$ the maximum
energy of electrons $E^e_{max}$ accelerated by shocks estimated in
the Bohm diffusion limit is
\begin{equation}
E^e_{max}\sim\sqrt{\frac{eHu^2}{3c\beta}} \,.
\end{equation}
that gives e.g. for the shock velocity $u=10^8$ cm s$^{-1}$, the magnetic
field strength $H=10^{-5}$ G and the energy density of relic
photons $w_{ph}=0.25$ eV cm$^{-3}$ the maximum energy of
accelerated electrons about $E^e_{max}\sim 5\times 10^{13}$eV.
\begin{figure}[ht]
\centering
\includegraphics[width=3.in]{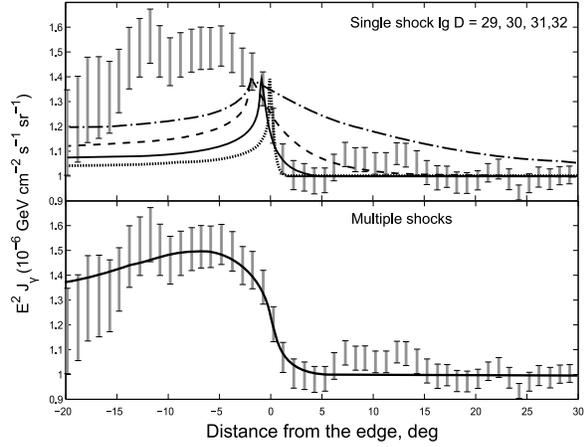} \caption{Spatial distribution of
the gamma-ray emission. Data are from \cite{meng}. {\it Top:} in case of single shock. Dotted line correspond to D=10$^{29}$ cm$^2$/s, solid to D=10$^{30}$ cm$^2$/s, dashed to D=10$^{31}$ cm$^2$/s and dash-dotted D=10$^{32}$ cm$^2$/s, {\it Bottom:} in case of several shocks distributed in accordance with (\ref{r_shock}).} \label{fig:spat}
\end{figure}
\begin{figure}[ht]
\centering
\includegraphics[width=3.in]{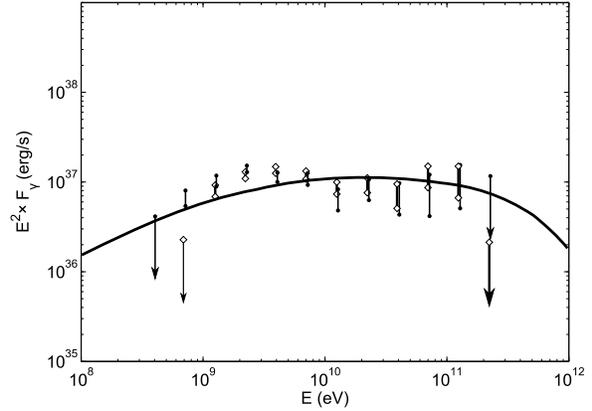} \caption{The spectrum of
gamma-ray emission from Fermi bubble in case of multi-shock
acceleration. Data points are taken from \cite{meng}.}
\label{fig:gamma}
\end{figure}

In Fig. \ref{fig:spat} we show the expected spatial distributions of
gamma-ray emission from the bubble for
  single shock  and  multiple
shocks cases.  From this figure one can
see that the single shock model  is unable to reproduce the  data. However,
for the parameters of star capture model these data are nicely
described.

The expected spectrum of gamma-ray emission from the Bubble due to
IC scattering of the electrons is shown in Fig. \ref{fig:gamma}.
\section{PROTON ACCELERATION IN THE BUBBLE AND THE ORIGIN OF THE "KNEE" COSMIC RAYS}
CRs within with energies below $E\sim 10^{15}$ eV  are generally attributed to SNRs in our
Galaxy. We assume that some of the CRs produced by SNRs in the Galactic disk are
re-accelerated by shocks in the Bubble to energy above $10^{15}$ eV that explains the origin of CRs beyond the "knee".

For the  multi-shock structure in the bubble an average distance $L$ between separate shocks given by
\begin{equation}
L=\tau_{0}u=30\left(\frac{\tau_0} {3\times 10^4\mbox{yr}}\right)\left(\frac{u}{10^8~cm/s}\right)pc.
\end{equation}
where $u$ is the shock front speed.

\begin{figure}
\centering
\includegraphics[width=3.in]{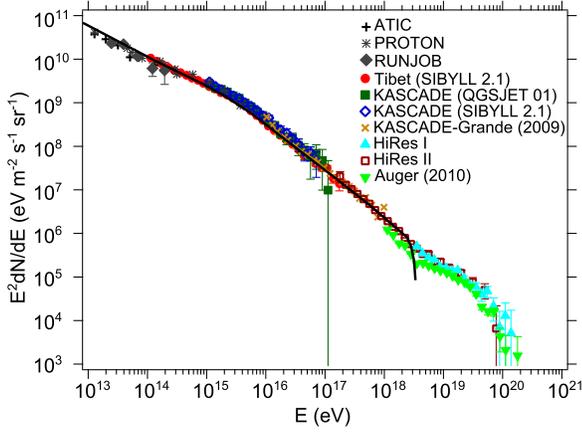} \caption{CR spectrum derived in the model of particle re-acceleration in the bubble. The data are summarized in \cite{kumiko}.} \label{fig:cr}
\end{figure}

If the value of $L$ exceeds the scale of particle acceleration by a single shock
which is $l_D\sim D/u$ where $D$ is the spatial diffusion
coefficient near a shock and $u$ is the  shock velocity, then particle acceleration by shocks is  pure stochastic, which describes by a momentum diffusion coefficient (see \cite{byk90})
\begin{equation}
\kappa\sim \frac{u^2}{cL}p^2 \label{kappa}\,.
\end{equation}
Then the equation describing particle production by SNRs in the disk, their re-acceleration in the bubble and propagation in the Galaxy can be presented in the form
\begin{eqnarray}
&&\frac{\partial}{\partial z}\left(D(\rho,p)\frac{\partial
f}{\partial z}\right) +\frac{1}{\rho}\frac{\partial}{\partial
\rho} \left(D(\rho,p)\rho\frac{\partial
f}{\partial\rho}\right)+\nonumber\\
&&+\frac{1}{p^2}\frac{\partial}{\partial p}\left(\kappa(\rho,p)
p^2\frac{\partial f}{\partial p}\right)=-Q(\rho,z,p)\,,
\label{dif_bubble}
\end{eqnarray}
where $\rho$ and $z$ are the cylindrical spatial coordinates, $p$
is the particle momentum. $D(\rho,p)$ is the spatial diffusion
coefficient, which is a function of coordinates and particle momentum, and
$Q(\rho,z,p)$ describes  CR injection by supernova
remnants in the disk with energies $E<10^{15}$ eV with the
spectrum $Q\propto p^{-4}$. The spectrum of CRs injected by SNRs and re-accelerated in the bubble is shown  in Fig. \ref{fig:cr}.
\section{Conclusion}
We have shown that series of shocks produced by a sequential
stellar captures by the central black hole can further
re-accelerate the protons emitted by SNRs up to energies above
$10^{15} eV$.  The predicted CR spectrum contributed by the Bubble
may be $E^{-\nu}$ where $\nu \sim 3$ for $10^{15}\mbox{ eV} < E <
10^{19}\mbox{ eV}$ that explains the knee CR spectrum.

The regime of electron acceleration in the bubble is quite
different from that of protons. It is a combination of single and
multishock accelerations. In this case we have a cut-off of the
electron spectrum at $E> 3~10^{13}$ eV and flattening of the
spectrum at $E<100$ GeV that explains nicely the bubble gamma-ray
spectrum and the sharp edge spatial distribution observed by
\cite{meng} if this emission is due to IC on the relic photons.

\bigskip 
\begin{acknowledgments}
 VAD and DOC are partly supported by the NSC-RFBR Joint Research
Project RP09N04 and 09-02-92000-HHC-a. KSC is supported by the GRF
Grants of the Government of the Hong Kong SAR under HKU 7011/10P.
CMK is supported, in part, by the Taiwan National Science Council
Grant NSC 98-2923-M-008-01-MY3 and NSC 99-2112-M-008-015-MY3. WHI
is supported by the Taiwan National Science Council Grant NSC
97-2112-M-008-011-MY3 and Taiwan Ministry of Education under the
Aim for Top University Program National Central University.
\end{acknowledgments}


\end{document}